\documentclass[lettersize,journal]{IEEEtran}
\usepackage{amsmath,amsfonts}
\usepackage{algorithmic}
\usepackage{algorithm}
\usepackage{array}
\usepackage[caption=false,font=normalsize,labelfont=sf,textfont=sf]{subfig}
\usepackage{textcomp}
\usepackage{amssymb}
\usepackage{stfloats}
\usepackage{url}
\usepackage{verbatim}
\usepackage{graphicx}
\usepackage{svg}
\usepackage{cite}
\usepackage{color}
\usepackage[normalem]{ulem}
\begin{document}
\setlength{\textfloatsep}{0.21cm}
\setlength{\abovedisplayskip}{0.1cm}
\setlength{\belowdisplayskip}{0.1cm}
\setlength{\baselineskip}{0.42cm}
\setlength{\intextsep}{1mm}
\setlength{\abovecaptionskip}{1mm}

\setlength{\textfloatsep}{2pt plus 1pt minus 1pt}
\setlength{\floatsep}{2pt plus 1pt minus 1pt}
\setlength{\intextsep}{2pt plus 1pt minus 1pt}

\setlength{\abovedisplayskip}{2pt plus 1pt minus 1pt}
\setlength{\belowdisplayskip}{2pt plus 1pt minus 1pt}
\setlength{\abovedisplayshortskip}{1pt plus 1pt minus 1pt}
\setlength{\belowdisplayshortskip}{1pt plus 1pt minus 1pt}

\setlength{\abovecaptionskip}{1pt}
\setlength{\belowcaptionskip}{-4pt}

\title{Pushing the Limits: Unlocking the Potential of Faster-than-Nyquist Signaling}

\author{Zichao~Zhang,~\IEEEmembership{Student Member,~IEEE,} 
        Melda~Yuksel,~\IEEEmembership{Senior Member,~IEEE,}
        Shuangyang~Li,~\IEEEmembership{Member,~IEEE,}
        Gökhan M. Güvensen,~\IEEEmembership{Member,~IEEE,}
Halim~Yanikomeroglu,~\IEEEmembership{Fellow,~IEEE}
\thanks{Z. Zhang and H. Yanikomeroglu are with the Department of Systems and Computer Engineering at Carleton University, Ottawa, ON, K1S 5B6, Canada; M. Yuksel and G. Guvensen are with the Department of Electrical and Electronics Engineering, Middle East Technical University, Ankara, 06800, Türkiye; Shuangyang Li is with the Technical University of Berlin, 10623 Berlin, Germany.}
%\thanks{Manuscript received October 1, 2024; revised December 1, 2024.}
}

% The article headers

%\IEEEpubid{0000--0000/00\$00.00~\copyright~2024 IEEE}

\maketitle

\begin{abstract}
%Faster-than-Nyquist (FTN) signaling is a non-orthogonal transmission technique that offers significant improvements in spectral efficiency for wireless communication systems. This article highlights the significant capacity gains achievable with FTN signaling by exploring its full range of acceleration factors, including very small values that result in considerable inter-symbol interference (ISI). We emphasize the importance of a clear definition of signal-to-noise ratio (SNR) in this context. Unlike Nyquist signaling—where performance comparisons under fixed transmit power or fixed received SNR yield consistent results—FTN performance can vary notably depending on which assumption is made. This distinction is particularly important when analyzing systems with varying acceleration factors, as small values can result in significant fluctuations in the instantaneous-to-average power ratio (IAPR) at the transmitter or lead to a reduction in the signal-to-noise ratio (SNR) per bit at the receiver. Based on these insights, we highlight application areas that can benefit from FTN transmission, reinforcing its relevance and the need for continued research in this area.
Faster-than-Nyquist (FTN) signaling is gaining attention as a smart way to pack more data into limited spectrum by intentionally breaking the traditional symbol-spacing rules. This article takes a fresh look at FTN’s potential to boost capacity, examining how performance varies across different acceleration factors and signal-to-noise ratio (SNR) definitions. Beyond the theory, we explore what it takes to make FTN work in practice, such as dealing with power amplifier constraints, managing high peak-to-average power, and designing practical coding strategies. We also highlight real-world issues like spectrum sharing, short-packet communication, and receiver complexity. With applications ranging from low-latency links to integrated sensing and satellite systems, FTN offers a compelling path forward for future wireless technologies.
\end{abstract}

\section{Introduction}

%\IEEEPARstart{M}{odern} communication systems typically adhere to the Nyquist criterion, transmitting consecutive data symbols in an orthogonal manner to avoid inter-symbol interference (ISI). This orthogonality ensures that signals do not overlap in time, allowing for straightforward symbol-by-symbol detection. However, in a seminal work dating back to 1975, Mazo challenged this belief by demonstrating that orthogonality is not a strict requirement for reliable communication \cite{mazo}. He showed that deliberately introducing ISI—by transmitting symbols faster than the Nyquist rate—could actually enhance spectral efficiency, without degrading the error performance significantly. Specifically, for binary phase shift keying (BPSK) with sinc pulses, sequence-based detection methods allowed up to a roughly 25\% increase in signaling rate without compromising error performance.

\IEEEPARstart{M}{odern} communication systems typically follow the Nyquist criterion, transmitting consecutive data symbols orthogonally to avoid inter-symbol interference (ISI). {This orthogonality enables straightforward symbol-by-symbol detection, even though the corresponding pulses may still overlap in time.} However, Mazo’s seminal 1975 work challenged this view, showing that orthogonality is not essential for reliable communication \cite{mazo}. By deliberately introducing ISI through {faster-than-Nyquist (FTN)} signaling, he demonstrated that spectral efficiency could be improved without significantly degrading error performance. Specifically, for BPSK with sinc pulses, sequence-based detection enabled up to a 25\% increase in signaling rate with no loss in reliability.

%Rusek et al. \cite{rusek} offered an early benchmark by evaluating the symbol-constrained capacity of FTN signaling, attributing its spectral-efficiency advantage to the exploitation of excess bandwidth. Their analysis, however, assumes the input symbols to be uniformly distributed and therefore does not achieve the unconstrained Shannon capacity. Adopting a spectral-domain viewpoint, Ganji et al. \cite{ganji} derived the FTN capacity in the frequency domain, where the capacity-achieving input distribution emerges naturally through spectrum-shaping. On the other hand, Kim \cite{kimproperty} and Takumi et al. \cite{takumi} approached the problem in the time domain by decomposing the FTN signal into an equivalent bank of parallel, interference-free sub-channels. This formulation enabled the derivation of an achievability scheme based on water-filling, which later turned out to be capacity achieving \cite{zhang2022faster}. Combining spatial multiplexing via MIMO with temporal packing via FTN presents a natural and promising direction for maximizing spectral efficiency as well. Recent studies, such as \cite{zhang2022faster} and \cite{zhang2024capacitypapranalysismimo} demonstrate that this combination can significantly enhance data rates. 

{
Rusek et al. \cite{rusek} provided an early benchmark for the symbol-constrained capacity of FTN, attributing its spectral-efficiency gains to excess-bandwidth utilization. However, their analysis assumes uniformly distributed inputs and therefore does not achieve unconstrained Shannon capacity. Ganji et al. \cite{ganji} later derived FTN capacity in the spectral domain, where the optimal input distribution emerges through spectrum shaping, while Kim \cite{kimproperty} and Takumi et al. \cite{takumi} studied the problem in the time domain by decomposing the FTN signal into parallel interference-free sub-channels. {This enabled an achievability scheme based on water-filling, which was later shown to be capacity achieving \cite{zhang2024capacitypapranalysismimo}. Combining MIMO spatial multiplexing with FTN temporal packing offers a natural and promising direction for maximizing spectral efficiency. It was shown in \cite{zhang2024capacitypapranalysismimo} that this integration can substantially boost data rates.}

Despite these potential gains, {FTN} signaling did not gain broad early adoption. Progress has been fragmented, with different studies addressing isolated aspects of FTN without forming a unified picture. This fragmented landscape calls for a more integrated understanding of FTN’s potential.

In this article, we first explain the trade-off between acceleration and capacity in MIMO-FTN systems over a wide range of acceleration factors. We then study instantaneous-to-average power ratio (IAPR) through the average outage probability of QPSK FTN signals and discuss its implications for transmitter design. We also review practical coded FTN implementations, receiver architectures, spectrum broadening, and spectrum-sharing issues, all of which are central to deployment. In particular, low-complexity receivers are important for reducing decoding cost, while spectrum broadening and multi-user sharing must be considered when assessing practical gains.

Building on this foundation, we further discuss the high-impact potential of FTN in emerging systems. In particular, we highlight its relevance in the finite block length regime for short-packet transmissions in hyper-reliable low-latency communications (HRLLC), as well as its promise for integrated sensing and communication (ISAC), where spectral efficiency and waveform flexibility are critical. These perspectives help connect theory with practice and outline a road map for FTN to evolve from a promising concept into a key enabler of next-generation wireless systems.
}

\begin{figure}[t]
    \centering
    \includegraphics[width=0.8\linewidth]{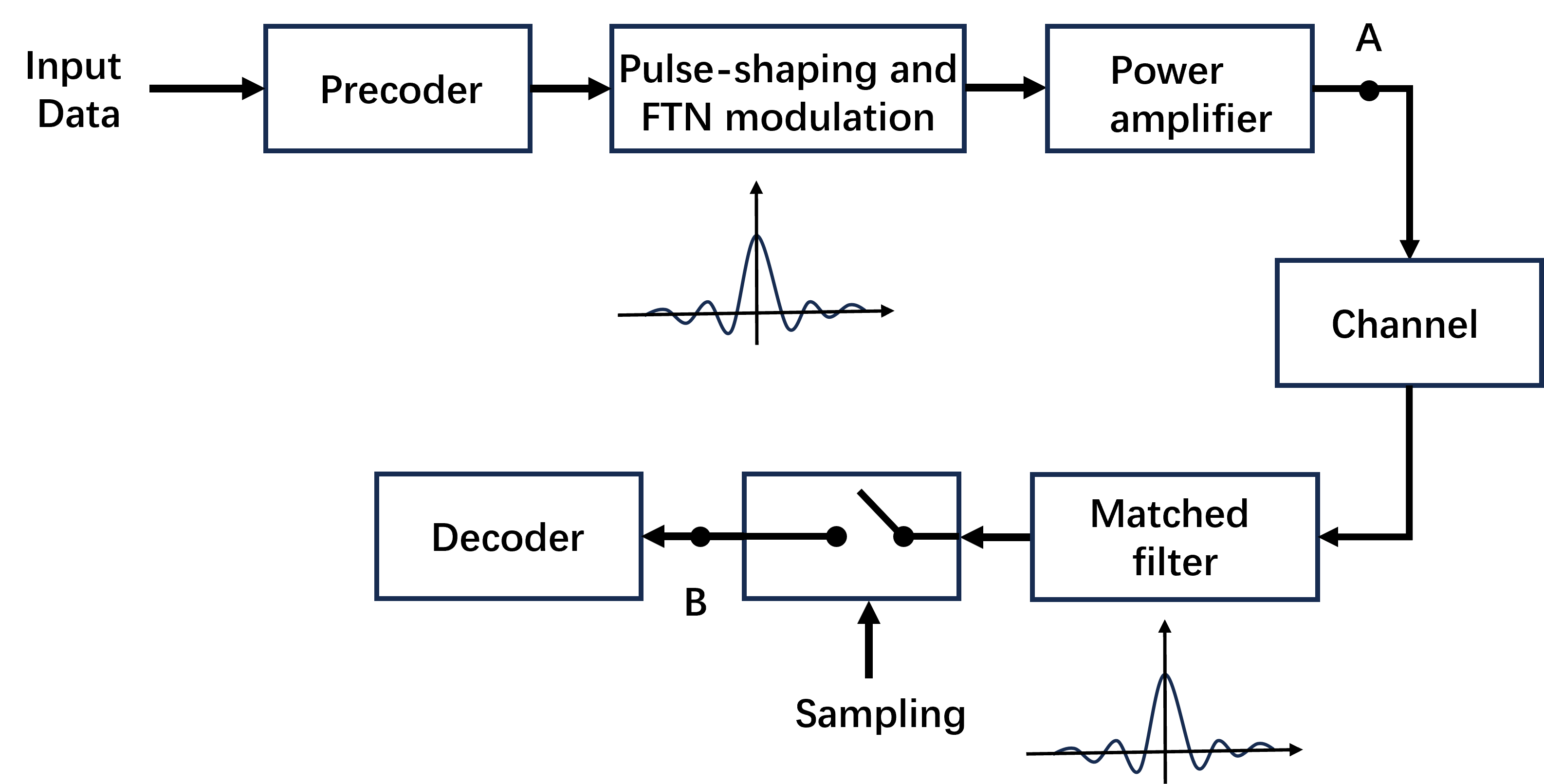}
    \caption{System model for FTN transmission. }
    \label{fig:sysmodel}
\end{figure}

\section{Information-Theoretic Foundations of FTN Signaling}
\label{sec:capftn}

The system model and the corresponding pulse sequence transmitted through the channel are illustrated in Figs.~\ref{fig:sysmodel} and~\ref{fig:isi}, respectively. In Fig.~\ref{fig:isi}, root-raised cosine (RRC) pulses with a roll-off factor of $\beta = 0.25$ are transmitted using an acceleration factor of $\delta = 0.8$ {and symbol period $T=1$}. Since $\delta \neq 1$, the value corresponding to Nyquist transmission, adjacent pulses are no longer orthogonal. As a result, symbols are placed closer together in time than the Nyquist criterion allows, and ISI is intentionally introduced. Although this interference may initially appear like a drawback, it can be carefully exploited to enhance spectral efficiency. However, this raises several fundamental questions: Can the acceleration factor be reduced arbitrarily close to zero? Does spectral efficiency always increase as the acceleration factor decreases? And how does the required transmit power evolve with varying acceleration? Addressing these questions requires a careful examination of how signal-to-noise ratio (SNR) is defined and interpreted in the FTN context.

\subsection{SNR Definitions in FTN}

In Nyquist transmission, the distinction between power and energy, or between received SNR and transmitted power, is typically not critical. Since the symbol duration is fixed, power and energy are linearly related, and thus the relative performance of transmission schemes remains unchanged regardless of the horizontal axis used in performance plots (e.g., bit error ratio (BER) vs. SNR). However, this equivalence does not hold in FTN signaling, where the acceleration factor $\delta$ alters the temporal spacing between symbols and introduces ISI. In this setting, careful attention must be paid to how SNR is defined, as it directly affects system behavior and performance.

Two main conventions exist in defining SNR for FTN systems: one assumes a fixed transmit power, while the other assumes a fixed received SNR. These choices have important implications. Under a fixed received SNR assumption, decreasing $\delta$ requires higher physical transmit power to maintain constant energy per symbol. This increase elevates the IAPR of the transmitted waveform. {As the symbol overlap becomes more pronounced with smaller $\delta$, the IAPR may exceed the linear operating range of power amplifiers, ultimately degrading system performance. }

\begin{figure}[t]
    \centering
    \includegraphics[width=0.9\linewidth]{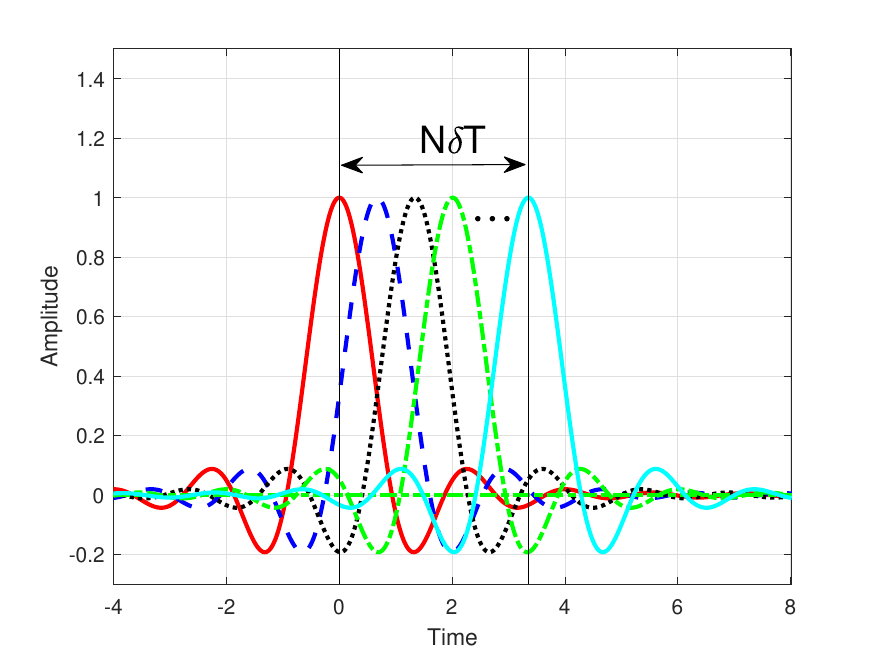}
    \caption{An example of ISI introduced by FTN, where $N$ symbols are transmitted with $T=1$, with acceleration factor $\delta=0.8$, and RRC pulses with roll-off factor $\beta=0.25$ are used.}
    \label{fig:isi}
\end{figure}

Formally, let $\sigma_0^2$ represent the power spectral density of additive white Gaussian noise. The transmit SNR is defined as $\mathsf{SNR_{tx}} = P/\sigma_0^2$, where $P$ is the average transmit power. {The receive SNR is given by $\mathsf{SNR_{rx}} = E/(T\sigma_0^2) = P\delta/\sigma_0^2$, where $E = P\delta T$ is the average energy per symbol.}  In Nyquist signaling ($\delta = 1$), the two definitions coincide. {Under FTN, however, they imply different behaviors: maintaining a constant transmit SNR holds transmission power $P$ fixed at the transmitter (e.g., point A in Fig.~\ref{fig:sysmodel}), but results in lower symbol energy and smaller Euclidean distances at the receiver (point B), increasing the error rate. Conversely, fixing receive SNR preserves constellation geometry at the detector but demands increasing transmit power as $\delta$ decreases in order to maintain symbol energy $E$, posing challenges for power amplifier linearity.} Additionally, these definitions assume asymptotic symbol lengths; for finite-duration transmissions, energy contributions from pulse tails (as shown in Fig.~\ref{fig:isi}) must also be considered.

{Both definitions are meaningful and relevant in different operating regimes. When the system operates at high transmit power, nonlinearities in the power amplifier become significant, making $\mathsf{SNR_{tx}}$ the limiting factor. In contrast, for battery-powered systems operating at low power, energy efficiency is critical, and $\mathsf{SNR_{rx}}$ becomes the primary constraint. In practice, both limitations may apply simultaneously, and system design must account for each.}

\vspace{-0.3cm}
\subsection{Capacity of FTN}

With the SNR definitions in place, we now examine how they influence the capacity of FTN signaling. Since the acceleration factor $\delta$ alters the symbol rate and energy per symbol, the choice between fixing transmit power or receive SNR leads to different capacity trends.

% In \cite{zhang2022faster} and \cite{zhang2024capacitypapranalysismimo}, the capacity of MIMO FTN is derived for arbitrary $\delta$. It is found that if $SNR_{tx}$ is fixed, for $\delta$ in the range $[\frac{1}{1+\beta}, 1]$, capacity increases with decreasing $\delta$. Once the threshold $\delta = \frac{1}{1+\beta}$ is reached, further acceleration does not improve capacity.

\begin{figure}[t]
    \centering
    \includegraphics[width=0.9\linewidth]{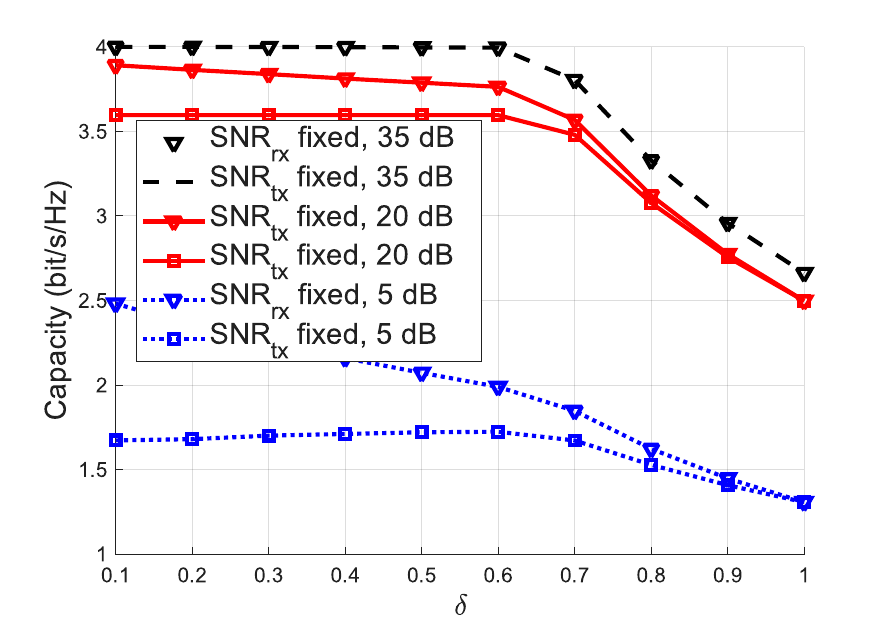}
    \caption{Constrained capacity vs $\delta$ for optimal power allocation scheme with different transmit SNR and received SNR values, where QPSK symbol set is used. The MIMO size is $2\times 2$. The pulse shape used is RRC with roll-off factor 0.5.}
    \label{fig:capacityvsdelta}
\end{figure}

In  \cite{zhang2024capacitypapranalysismimo}, the capacities of FTN for both frequency-flat and frequency-selective MIMO channels are derived for arbitrary $\delta$. It is found that if $\mathsf{SNR_{tx}}$ is fixed, for $\delta$ in the range $[\delta_{th}, 1]$, where $\delta_{th}$ is a pulse shape dependent threshold acceleration factor, capacity increases with decreasing $\delta$. Once the threshold $\delta_{th}$ is reached, further acceleration does not improve capacity. Although more symbols are transmitted per second, each symbol observes more severe ISI, limiting the achievable information rate. On the other hand, if $\mathsf{SNR_{rx}}$ is fixed, capacity continues to improve with faster signaling, since constellation geometry is preserved at the output of the matched filter shown in Fig.~\ref{fig:sysmodel}. 
{This behavior is revisited in Fig.~\ref{fig:capacityvsdelta} for $2 \times 2$ frequency-flat MIMO system, where the MIMO channel coefficients are i.i.d. and complex Gaussian distributed according to  $\mathcal{CN}\left(0,\frac{1}{2}\right)$. In particular, we now report constrained capacity results with practical finite-alphabet signaling (QPSK). Fig.~\ref{fig:capacityvsdelta} shows that under fixed $\mathsf{SNR_{rx}}$, we observe that at $\mathsf{SNR_{rx}}=5$~dB, the constrained capacity increases as $\delta$ decreases over the considered range. At higher SNR (namely, 35~dB), the gain saturates due to the finite-alphabet limitation, so the improvement is not unbounded. Therefore, we understand that, higher symbol rate can still provide a clear rate benefit in the low-to-moderate SNR regime for fixed $\mathsf{SNR_{rx}}$.} 
Note that, capacity of FTN for acceleration factor $0$ is undefined as the symbols become indistinguishable under excessive overlap.

The high-level principle behind the capacity achieving transmission strategy for MIMO FTN is quite intuitive. For frequency-flat channels, the optimal scheme separates spatial and temporal processing: the MIMO channel is first decomposed into parallel eigen-channels using singular value decomposition, and water-filling is applied to allocate power across them. Then, FTN precoding is applied in the time domain for each eigen-channel to mitigate ISI. This is equivalent to spectral shaping in the frequency domain. In frequency-selective channels, the approach is similar, but there are now two sources of ISI: channel-induced multi-path fading and FTN-induced overlap. The spectral shaping must therefore address both simultaneously. 

\subsection{Optimal vs. Uniform Power Allocation}

FTN systems benefit from joint power allocation in both the spatial and frequency domains. Spatially, water-filling distributes power across MIMO eigen-channels. Spectrally, inverse shaping allocates power across frequencies to counter ISI. These allocations are independent and can be combined flexibly. To reap the benefits of FTN signaling, optimal power allocation both in time and frequency is not necessary. For example, uniform power allocation both across space and time offers a suboptimal, yet practical design. %As illustrated in Fig.~\ref{fig:capacityvsdelta}, FTN signaling achieves significantly higher spectral efficiency than Nyquist transmission, even when uniform power allocation (both in space and frequency) is employed.

\begin{figure}[t]
    \centering
    \includegraphics[width=0.9\linewidth]{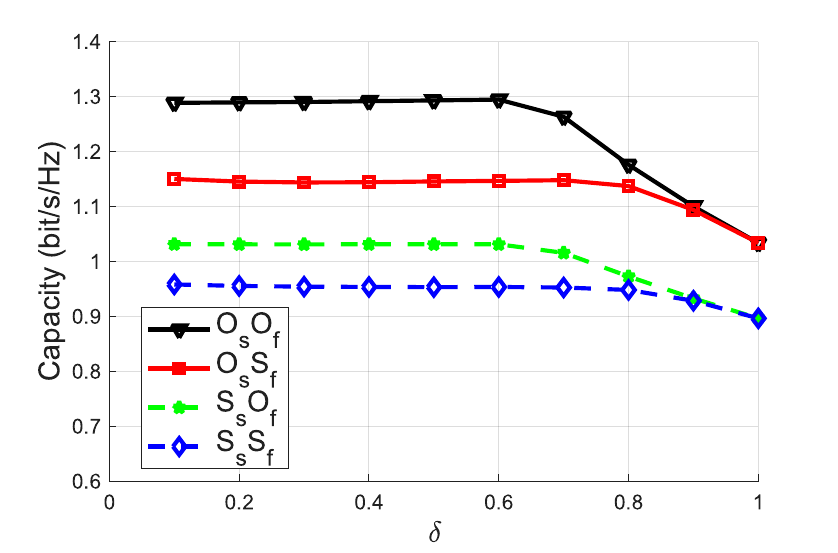}
    \caption{Constrained capacity vs $\delta$ for four different power allocation schemes with transmit SNR $\mathsf{SNR_{tx}}=5 $ dB where QPSK symbol set is used. The MIMO size is $2\times 2$. The pulse shape used is RRC with roll-off factor 0.5. In the legend, $O_s$ and $S_s$ respectively mean optimal and suboptimal in space. Similarly, $O_f$ and $S_f$ respectively mean optimal and suboptimal in frequency.}
    \label{fig:tauvsctxsnr}
\end{figure}

{Fig.~\ref{fig:tauvsctxsnr} offers a clearer perspective on the impact of different power allocation strategies on the constrained capacity where the channel configuration is the same as in Fig.~\ref{fig:capacityvsdelta}.} It reveals that optimal power allocation across transmit antennas; i.e., spatial power allocation, has a greater impact on performance than frequency-domain precoding used to mitigate ISI in FTN. This is an encouraging result for FTN systems, as it suggests that high performance can be achieved without relying heavily on transmitter-side complexity for precoding. 
%In addition, it is well known that at high SNR, power allocation among transmit antennas becomes less critical, further reducing the need for sophisticated transmitter adaptation.

%\subsection{Looking Ahead}

These theoretical findings offer useful guidelines for FTN system design. Modulation and coding can be tailored based on whether transmit or receive SNR is constrained. The threshold acceleration factor $\delta_{th}$, which is equal to $1/(1+\beta)$ in the case of RRC pulses with roll-off factor $\beta$, serves as a practical design rule: reducing $\delta$ beyond this point yields no capacity returns in case of transmit power limitations.

%{The results in Section~II characterize the potential spectral-efficiency gains of FTN under different SNR conventions and power-allocation strategies. However, translating these theoretical gains into practice critically depends on the transmitter’s ability to radiate the FTN waveform under realistic power-amplifier (PA) constraints. In particular, as the acceleration factor $\delta$ decreases, the increased pulse overlap can produce large envelope fluctuations, raising the instantaneous-to-average power ratio (IAPR). This in turn necessitates PA power back-off to maintain linear operation, effectively reducing the usable transmit power (and hence the effective received SNR), which may significantly diminish the gains predicted by capacity-oriented analysis. Motivated by this practical bottleneck, we next study the IAPR/outage behavior of FTN waveforms and its implications for transmitter design.}

\section{Instantaneous to Average Power Ratio Analysis on FTN}
\label{sec:iaprftn}

{
In FTN transmission, stronger acceleration packs pulses more tightly, increasing overlap and thus IAPR. Because a transmitter power amplifier has limited dynamic range (saturation), maintaining linear operation requires reducing input power to accommodate peaks (\emph{power back-off}, in dB). Without compensation, the needed back-off is approximately the IAPR, i.e., the ratio of instantaneous to average power.

Based on the IAPR distribution, the probability that instantaneous power exceeds the back-off threshold—called outage probability—can be computed. For cyclostationary pulse trains, we examine the average outage probability: the average chance that instantaneous power exceeds the threshold over one period. {This provides a time-averaged characterization of the transmitted power behavior.} The average outage probability depends strongly on whether $\mathsf{SNR_{tx}}$ or $\mathsf{SNR_{rx}}$ is fixed, and also on the modulation constellation.

With Gaussian signaling (capacity-achieving), the transmitted pulse train is a Gaussian process. If $\mathsf{SNR_{tx}}$ is fixed, decreasing the acceleration factor does not change the process distribution (since physical transmit power is fixed), so the average outage probability is unchanged with acceleration. If $\mathsf{SNR_{rx}}$ is fixed, the average outage probability increases as $\delta$ decreases and eventually approaches a constant---meaning instantaneous power takes larger values as $\delta$ decreases, and in the limit becomes infinite almost surely.
}

{
{Although Gaussian signaling is useful for theoretical analysis, practical FTN systems must also be evaluated with finite constellations such as PSK and QAM.}  Fig.~\ref{fig:qpskrxsnr} shows the single-antenna FTN results for $\mathsf{SNR_{tx}}=20$~dB and $\mathsf{SNR_{rx}}=20$~dB. {With QPSK, the average outage probability increases as $\delta$ decreases in both the $\mathsf{SNR_{tx}}$-fixed and $\mathsf{SNR_{rx}}$-fixed cases.}  {Here, $\gamma$ denotes the IAPR (power back-off) threshold, so the outage probability is defined as
$P_{\mathrm{out}}(\gamma)\triangleq \Pr\{\mathrm{IAPR}>\gamma\}$,
i.e., the probability that the instantaneous transmit power exceeds $\gamma$ times the average power
(equivalently, exceeds a $\gamma$~dB back-off limit).}
{Fig.~\ref{fig:qpskrxsnr} shows that, under practical finite constellations, the required  IAPR/power back-off is influenced more strongly by modulation order than by the FTN acceleration factor $\delta$ in the tested range. In particular, while higher-order constellations require visibly larger back-off, the additional penalty caused by decreasing $\delta$ remains relatively moderate in the fixed-$\mathsf{SNR_{tx}}$ case, suggesting that FTN does not necessarily impose a severe IAPR disadvantage and should instead be evaluated jointly with modulation order and the adopted SNR definition.}

\begin{figure}[t]
    \centering
    \includegraphics[width=0.9\linewidth]{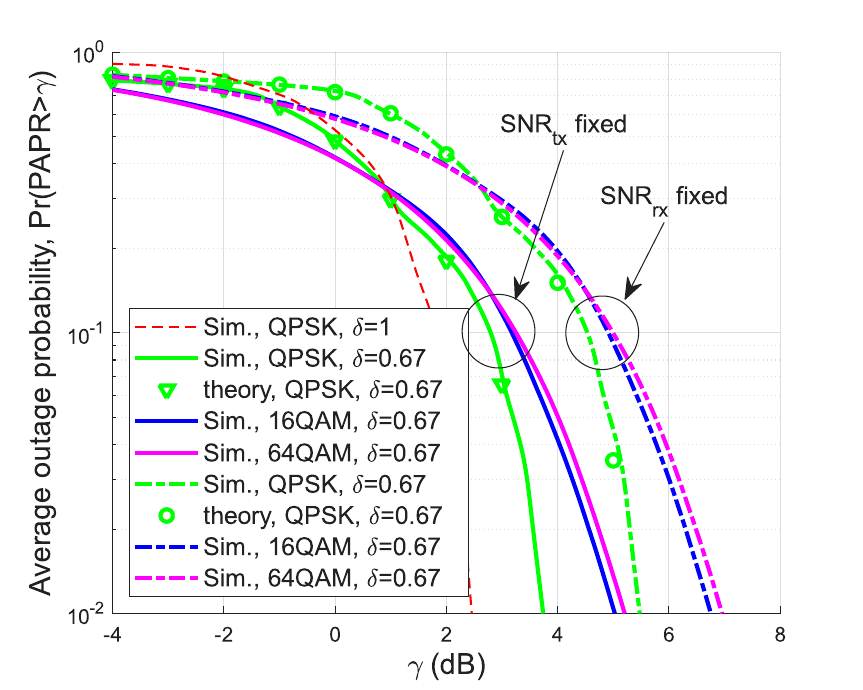}
    \caption{Average outage probability of single-antenna FTN versus $\delta$ for uniform power allocation in space and frequency with $\beta=0.5$.  The theoretical curve is also shown.}
    \label{fig:qpskrxsnr}
\end{figure}

%As $\delta$ decreases, the IAPR distribution degrades mildly when $\mathsf{SNR_{tx}}$ is fixed, but the degradation is more pronounced when $\mathsf{SNR_{rx}}$ is fixed.
Intuitively, in the $\mathsf{SNR_{rx}}$-fixed case the symbol energy remains constant across $\delta$, whereas in the $\mathsf{SNR_{tx}}$-fixed case the symbol energy effectively decreases with $\delta$ because the physical transmit power is held constant. As a result, the power amplifier (PA) output power scales as $E/(\delta T)$ under $\mathsf{SNR_{rx}}$ fixation, but stays proportional to $P$ under $\mathsf{SNR_{tx}}$ fixation. For the same threshold $\gamma$, the $\mathsf{SNR_{rx}}$-fixed setting is therefore more likely to exceed $\gamma$, so its outage curve lies above the $\mathsf{SNR_{tx}}$-fixed curve and approaches its limiting value faster as $\delta\to 0$. This highlights that preserving symbol energy under faster-than-Nyquist signaling can significantly worsen IAPR, so IAPR should be treated as a practical penalty when comparing BER: although fixing $\mathsf{SNR_{rx}}$ increases the effective Euclidean distance, it also demands substantially higher PA output power.

More broadly, IAPR depends on constellation design, pulse shape, and acceleration. ISI is weighted by sampled pulse values and is dominated by nearby symbols; for RRC pulses, a larger roll-off $\beta$ produces smoother waveforms, reduces overlap, and typically improves IAPR. As $\delta$ decreases, overlap intensifies and large peaks become more likely. Constellations with larger amplitude variations (e.g., QAM, especially outer points in 16-QAM) tend to exhibit worse IAPR, whereas PSK---having constant amplitude---typically performs better. These considerations are essential in practical FTN design.
}

\section{Coded FTN Signaling}
\label{sec:codeftn}

%{In practical FTN systems, channel codes are often employed to cope with ISI. From a communication theoretical perspective, the application of channel codes can significantly improve the minimum Euclidean distance among different FTN codewords, leading to an improved error performance. For coded FTN systems, receivers are usually designed to follow the structure of Turbo equalization, in which the outputs of the FTN detector and the channel decoder are iteratively updated by exchanging extrinsic information. However, most of the existing channel codes are designed for additive white Gaussian noise (AWGN) channels, where no iterative processing is needed. Therefore, directly adopting such codes for FTN can lead to degraded error performance.

{In practical FTN systems, channel codes are often employed to cope with ISI. From a communication-theoretic perspective, the application of channel codes can improve the minimum Euclidean distance among different FTN codewords, leading to improved error performance. For coded FTN systems, the receiver is often designed according to the principle of Turbo equalization, where the FTN detector and the channel decoder iteratively exchange extrinsic information. Many standard channel codes, however, are originally designed and optimized for ISI-free channels, such as additive white Gaussian noise channels. When these codes are applied to FTN signaling, the achievable performance depends on the specific iterative receiver design, including the FTN detector, the exchanged extrinsic information, and the channel decoder. Therefore, careful receiver design is important for fully exploiting the benefit of coded FTN signaling.}

In~\cite{yang2025JSAC}, a family of tailored low-density parity-check (LDPC) codes is proposed for FTN signaling. These optimized LDPC codes are obtained by applying a masking operation to the base matrix of the standard 5G LDPC codes, aiming to achieve a lower decoding threshold and thereby improved decoding performance for coded FTN signaling under a given shaping pulse and $\delta$. Particularly, the raptor-like structure and rate compatibility are preserved in the proposed LDPC codes, and the encoder and decoder are reused with only minor modifications.

\begin{figure}[t]
    \centering
    \includegraphics[width=0.9\linewidth]{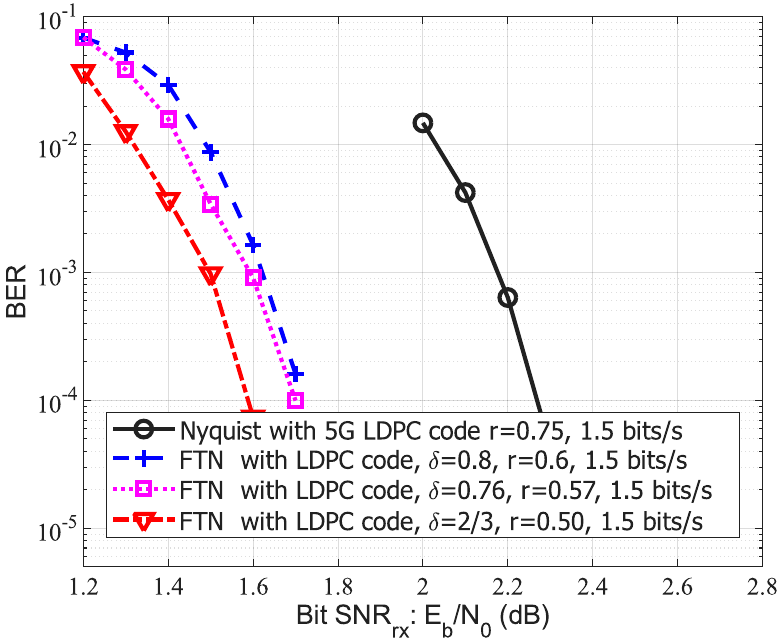}
    \caption{The error performance comparison between coded FTN system and coded Nyquist system, where the results are compared under roughly the same spectral efficiency. The code rate $r$ is also given. }
    \label{fig:codedFTN}
\end{figure}

The error performance of coded FTN system is compared with the Nyquist counterpart in Fig.~\ref{fig:codedFTN}. In the simulation, both FTN and Nyquist transmissions apply the RRC pulse with a roll-off factor $\beta=0.3$. Moreover, $\mathsf{SNR_{rx}}$ is fixed for a finite number of symbols; thus, the tail behavior of the RRC pulses is also taken into account. In the figure, the coded FTN system applies the optimized LDPC codes with code rates $r=0.6$, $0.57$, and $0.5$ from~\cite{yang2025JSAC}, corresponding to $\delta\in \{0.8,0.76,\frac{2}{3}\}$, while the coded Nyquist systems consider the original 5G LDPC codes. For a fair comparison, all the codeword lengths in the simulation are around $11000$, and the corresponding spectral efficiencies are roughly the same, where BPSK mapping is considered. From the figure, we observe that the optimized LDPC-coded FTN system achieves roughly $0.6$, $0.6$, and $0.7$~dB SNR gain compared to the Nyquist LDPC-coded system for different $\delta$ values, under the same spectral efficiency. Furthermore, we also notice that the error performance of the coded FTN system can be further improved when $\delta$ decreases, which is reasonable because for smaller $\delta$, a lower-rate LDPC code with stronger error-correction performance is adopted to achieve the same spectral efficiency.

%From the figure, we observe that the coded FTN system enjoys roughly $0.7$ dB and $1$ dB SNR gain compared to the Nyquist LDPC code and the Nyquist Turbo code, under the same spectral efficiency, where it requires less SNR compared to the constrained capacity of Nyquist signaling with BPSK constellations. Furthermore, we also notice that the error performance of the coded FTN system is still roughly $0.75$ dB away from the FTN Shannon limit, indicating the great potential of coded FTN signaling.

\section{Implementation Challenges and Practical Considerations}

{FTN signaling shows strong potential for improving spectral
efficiency, but its practical deployment still faces several
implementation challenges.} In practical systems, challenges like finite-block length transmission, increased complexity in precoding, equalization and decoding, and the need to work with imperfect channel state information all come into play. Precoding for FTN can lead to a spectrum-broadening side effect. It can also put pressure on synchronization and spectrum sharing. {Efficient hardware implementation also remains an important issue.
These concerns must be addressed before FTN can be widely deployed.}

{\subsection{Complexity of FTN Transmission}
\label{sec:ftncomplexity}

In practical FTN receivers, complexity is primarily driven by three factors: \emph{effective memory growth}, \emph{filtering/oversampling burden}, and \emph{iteration-driven latency}. As the acceleration factor decreases, the effective ISI memory length increases, which leads to exponentially increased trellis size for trellis/sequence-based detectors (e.g., MLSE/BCJR).  Reduced-memory families such as truncated-memory MAP/RSSE/DFE type structures explicitly limit the memory length to yield a controllable complexity and enable a tunable performance--complexity trade-off. In contrast, linear equalization such as LMMSE in time or frequency domain typically scales polynomially with filter length or block size for direct block solvers, with FFT-based implementations making it more compatible with real-time platforms. Finally, when iterative detection/decoding is used, total runtime and latency scale approximately linearly with the number of iterations, so the iteration count becomes a key practical constraint in low-latency or power-limited deployments. {A detailed runtime benchmarking of BCJR and LMMSE receivers under different acceleration factors requires an implementation-dependent study, involving memory truncation, modulation order, channel length, and stopping/iteration rules.}
}

% \subsection{Finite Block Length Transmission}

% The theoretical capacity results shown in this article in Figs.~\ref{fig:capacityvsdelta} and~\ref{fig:tauvsctxsnr} are achievable only if the {block length} is infinity. However, this assumption is impractical for real-world applications, particularly in HRLLC scenarios, where finite {block lengths} are required. Therefore, it is essential to examine the feasibility of FTN signaling under finite {block length} constraints.  

% In \cite{zhangfblftn}, a comparative analysis between Nyquist and FTN signaling in finite block length is conducted. It is demonstrated that with $\mathsf{SNR_{tx}}$ fixed, an FTN system with an acceleration factor $\delta = 0.85$ and a block length of $N = 200$ achieves the same channel coding rate and BER as the Nyquist system with a block length of $N = 2000$. 
% {\color{blue}Here, the channel coding rate refers to the maximum code rate considered in the finite-blocklength analysis, rather than the total number of transmitted information bits. Therefore, this comparison indicates that the same code-rate can be achieved with a shorter FTN block under the considered setting, which makes it relevant for latency-sensitive finite-blocklength transmission scenarios.}

{\subsection{Finite-Blocklength and Latency-Sensitive FTN Transmission}

The theoretical capacity results shown in this article in Figs.~\ref{fig:capacityvsdelta} and~\ref{fig:tauvsctxsnr} are achievable only when the block length tends to infinity. However, this assumption is impractical for real-world applications, particularly in latency-sensitive scenarios such as HRLLC, where finite block lengths are required. Therefore, it is essential to examine the feasibility of FTN signaling under finite-blocklength constraints.

In \cite{zhangfblftn}, a comparative analysis between Nyquist and FTN signaling in finite block length is conducted. It is demonstrated that with $\mathsf{SNR_{tx}}$ fixed, an FTN system with an acceleration factor $\delta = 0.85$ and a block length of $N = 200$ achieves the same finite-blocklength spectral efficiency  as the Nyquist system with a block length of $N = 2000$. Here,  the finite-blocklength spectral efficiency refers to the maximum code rate considered in the finite-blocklength analysis, rather than the total number of transmitted information bits. Namely, the payload of these two cases are still different. Therefore, this comparison indicates that comparable finite-blocklength spectral
efficiency can be achieved with a shorter FTN block under the considered
setting, which makes it relevant for latency-sensitive finite-blocklength
transmission scenarios.

Due to the inherent time-packing structure of FTN signaling, data symbols are transmitted with spacing $\delta T$ instead of the Nyquist spacing $T$. Hence, for the same number of transmitted symbols, FTN occupies a shorter signaling interval than Nyquist transmission, up to pulse-tail and implementation overheads. This characteristic makes FTN particularly attractive for time-sensitive applications, such as drone and vehicle communications, where reducing transmission duration while maintaining spectral efficiency is important for real-time data exchange. Nevertheless, the actual achievable latency gain also depends on receiver complexity, equalization, decoding latency, and implementation constraints, as discussed in Section~\ref{sec:ftncomplexity}.}

\subsection{Inter-Spectrum Interference}

The capacity enhancement in FTN signaling stems from exploiting excess bandwidth by increasing the signaling rate \cite{rusek}. This excess bandwidth is a result of using smoother pulses for symbol modulation. However, the spectral tails associated with excess bandwidth can lead to interference between adjacent frequency bands. In conventional systems, weak interference is typically treated as noise and discarded with minimal impact on performance. In contrast, FTN signaling leverages excess bandwidth to improve the information rate, making spectrum truncation more detrimental than in Nyquist signaling. As noted in \cite{spectrumsharing}, if among multiple channels, spectrum is overlapped and while decoding treated as noise, FTN gains can be limited. Therefore, advanced signal processing techniques should be studied to eliminate any potential inter-spectral interference that FTN may result. %These advanced techniques are likely to be based on multiple access channel principles as proposed in \cite{amacftn}.

\subsection{Spectrum Broadening}

% In practical implementations, it is neither feasible nor desirable to use pulses with infinite duration or infinite bandwidth. Instead, approximately time-limited and approximately bandwidth-limited pulse shapes are preferred. When a theoretically time-infinite pulse is truncated for transmission, this is equivalent to multiplying the pulse by a rectangular window in the time domain. In the frequency domain, this truncation results in the pulse spectrum being convolved with a sinc-shaped spectrum, introducing spectral ripples beyond the desired bandwidth. These ripples are typically negligible in magnitude compared to the main spectrum and can often be disregarded and an approximately time-limited and approximately bandwidth-limited pulse shape is obtained.  

% In FTN signaling, if the transmitted symbols are independent and identically distributed (i.i.d.), the data spectrum remains flat, meaning FTN alone does not alter the spectral characteristics and has the same spectral bandwidth as Nyquist transmission. However, when optimal precoding is applied for time-truncated pulses described in the paragraph above, the spectrum shaping introduced by the precoding process interacts with the inherent spectral ripples of the pulse and amplifies them. As a result, these magnified ripples effectively expand the occupied bandwidth, increasing the system’s effective bandwidth beyond that of conventional Nyquist signaling. Therefore, there is a tradeoff between precoding gains and spectral bandwidth in FTN.

{
In practice, pulses cannot be truly time-infinite or strictly band limited, so systems use waveforms that are only approximately time and bandwidth limited. Truncating an ideal time-infinite pulse is equivalent to multiplying it by a rectangular window in time, which in frequency corresponds to convolving the pulse spectrum with a sinc function. This creates small spectral ripples outside the nominal band; they are usually much weaker than the main lobe and can often be neglected, yielding an approximately time and band-limited pulse.

For FTN with i.i.d.\ symbols, the data spectrum is flat, so FTN itself does not change the occupied bandwidth relative to Nyquist signaling. However, when optimal precoding is applied to compensate for time-truncated pulses, the resulting spectral shaping can interact with these ripple components and amplify them. The magnified ripples effectively increase the occupied bandwidth, so FTN precoding can trade performance gains for additional bandwidth expansion.}

\subsection{Practical Advances in FTN Implementations and Prototypes}

%While FTN traditionally depends on complex sequence estimation to handle inter-symbol interference, recent studies show that even simple, short-memory detection techniques can recover much of the performance gain, making FTN more practical than it once seemed \cite{ebrahim}. That said, there are still several real-world challenges ahead. Techniques like precoding, equalization, and decoding can get computationally heavy \cite{1_R1}, especially when systems need to run in real time or on power-limited hardware.  %{On top of that, FTN performance under imperfect channel information or timing errors are open issues \cite{imperfect}.} Overcoming these challenges is key to making FTN a viable option in future communication systems, from dense 6G networks to compact satellite links.

{While FTN signaling can provide spectral-efficiency gains by intentionally introducing inter-symbol interference, its practical implementation still requires efficient methods for handling the resulting ISI. In this regard, techniques such as precoding, equalization, and decoding can get computationally heavy \cite{1_R1}, especially when systems need to run in real time or on power-limited hardware. In parallel with these algorithmic developments, recent experimental and prototyping results have begun to validate FTN feasibility beyond simulation. For example, an FTN multi-band carrierless amplitude and phase (CAP) visible-light communication prototype reports a 50\% compression ratio and an average achievable spectral efficiency exceeding 6.34~bit/s/Hz under a $3.8\times 10^{-3}$ BER threshold, highlighting practical operation under bandwidth-limited devices and FTN-induced interference \cite{ftncap}.  Moreover, a proof-of-concept FTN ``super-signal'' experiment demonstrates 270~GBd FTN transmission within a 252.4~GHz bandwidth, illustrating end-to-end FTN waveform generation and transmission in an experimental setting with low-speed electronics and photonics \cite{natureftn}.  In addition, hardware-oriented multi-carrier FTN studies show that FTN can be incorporated into FFT/IFFT-based transceivers with orthogonal/FTN mode switching \cite{rusekhardware}. Fixed-point FPGA/CMOS results close to floating-point performance are also demonstrated. These results provide further evidence of FTN’s practical feasibility, while also reinforcing the need for complexity-aware receiver design and hardware-conscious waveform optimization.}

\section{Future Directions and Research Opportunities}

The above foundation enables informed deployment of FTN signaling in real-world systems. In the remainder of this article, we explore how these insights guide FTN applications in 
HRLLC and ISAC systems. We also discuss the potential of applying FTN in the delay-Doppler domain and also in satellite communications.

\subsection{FTN Integration Considerations in OFDM-based 5G/6G Systems}

%FTN is often introduced in a single-carrier setting, whereas 5G NR (and many 6G candidates) are largely OFDM-based. It is necessary to investigate the coexistence and synergy between FTN and OFDM. Specifically, consider an OFDM system with $N$ subcarriers where only $N_c$ subcarriers are active (the remaining $N-N_c$ are nulled), which corresponds to frequency-domain zero padding and generating $N$ time-domain samples per OFDM symbol. We then transmit these $N$ samples using FTN with spacing $\delta T$. When $\delta$ is below the threshold, spectrum folding from FTN creates spectral regions that are identically zero {\color{blue}\cite{zhang2024capacitypapranalysismimo}}; by matching the OFDM occupied band to the 
%{\color{purple}:clarify the non-zero region}
%non-zero region, we obtain the relationship 
%$\frac{N_c}{N}=\frac{(1+\beta)/T}{1/(\delta T)}=\delta(1+\beta)$, 
%which links the active-subcarrier ratio to the FTN acceleration factor $\delta$. {Here, the non-zero region means the usable spectral support after folding, i.e., the frequency interval(s) where the folded FTN spectrum still carries signal energy.} Under this construction, FTN-induced interference is avoided by design, so a standard OFDM receiver can be used without additional FTN equalization while still achieving a shorter effective symbol time.

{FTN is often introduced in a single-carrier setting, whereas
5G NR (and many 6G candidates) are largely OFDM-based.
It is necessary to investigate the coexistence and synergy
between FTN and OFDM. Specifically, consider an OFDM
system with $N$ subcarriers where only $N_c$ subcarriers are
active (the remaining $N-N_c$ are nulled), which corresponds
to frequency-domain zero padding and generating $N$ time-domain
samples per OFDM symbol. These samples can be regarded as the
symbol sequence to be transmitted through the subsequent FTN
pulse-shaped waveform, where the same shaping pulse is used and
the symbol spacing is reduced from $T$ to $\delta T$. Equivalently, if one OFDM symbol has duration \(T_{\rm OFDM}\), the
conventional IDFT output samples have chip/sample interval
\(T_{\rm OFDM}/N\), while the combined FTN-OFDM construction uses a
faster chip/sample interval \(\delta T_{\rm OFDM}/N\). Thus, the chip
rate is increased from \(N/T_{\rm OFDM}\) to \(N/(\delta T_{\rm OFDM})\). In this sense,
the acceleration is introduced in continuous time through pulse-shaped
transmission, rather than by compressing a fixed OFDM waveform, and
the resulting non-orthogonality comes from the overlap between
neighboring pulse-shaped symbols. When $\delta$ is below the
threshold, spectrum folding from FTN creates spectral regions that are
identically zero~\cite{zhang2024capacitypapranalysismimo}. By matching the OFDM
occupied band to the non-zero region, we obtain the relationship
$N_c/N=\delta(1+\beta)$, which links the active-subcarrier ratio to
the FTN acceleration factor $\delta$. Here, the non-zero region means
the usable spectral support after folding, i.e., the frequency
interval(s) where the folded FTN spectrum still carries signal energy.
Under this construction, FTN-induced interference is avoided by design,
so a standard OFDM receiver can be used without additional FTN
equalization while still achieving a shorter effective symbol time.}

%\subsection{HRLLC}

%Due to the inherent structure of FTN signaling, data symbols are transmitted within a shorter time interval compared to Nyquist signaling, effectively reducing the overall communication duration. This characteristic makes FTN particularly advantageous for time-sensitive applications, such as drone and vehicle communications, where minimizing latency and maximizing spectral efficiency are critical for real-time data transmission. However, the actual achievable latency depends on the receiver complexity as discussed in Section~\ref{sec:ftncomplexity}.

\subsection{ISAC}

ISAC is an emerging technology expected to play a key role in future wireless systems. By integrating sensing into communication, ISAC not only enhances data transmission but also leverages sensing outcomes to optimize performance. Applications span autonomous vehicles and indoor Wi-Fi networks. FTN improves efficiency by compressing symbols into tighter time frames. When ISI is properly managed, FTN can boost both communication and sensing rates in ISAC systems \cite{ftnisac}.

{FTN is well-suited for integration with ISAC waveforms to
offset rate losses from sensing. {However, ISAC waveforms can be particularly sensitive to PA nonlinear
distortion, since such distortion may degrade both communication quality and
sensing accuracy.}
Recent studies have therefore considered ISAC waveform design
with adjustable IAPR for PA-related concerns~\cite{ref02rev5}.
Similarly, in FTN-enabled ISAC, the acceleration factor and
transmit power should be chosen with PA linearity and back-off
constraints in mind.}

\subsection{Delay-Doppler Domain FTN Signaling}

%OTFS, and in general the delay-Doppler (DD) waveform, have recently been acknowledged as an effective means to combat double selectivity in wireless channels. Different from the conventional symbol multiplexing in either the time or frequency domain, the DD waveform considers symbol multiplexing in the DD domain, where an effective DD shaping pulse is in place to carry information symbols. Particularly, due to limited time and frequency resources, the effective DD shaping pulse exhibits certain excess occupancy in delay and Doppler resources while roughly maintaining orthogonality among different DD information symbols. Therefore, the consideration of FTN signaling in the DD domain is well-motivated, as it can effectively collect the additional degrees of freedom due to this excess resource occupancy, at the expense of increased equalization complexity. Both theoretical and algorithmic studies on DD-domain FTN signaling are still in their infancy, especially under complex doubly selective channels, and further investigation in this direction requires attention.

{OTFS, and in general the delay-Doppler (DD) waveform, have recently been acknowledged as an effective means to combat double selectivity in wireless channels. Different from conventional symbol multiplexing in either the time or frequency domain, DD waveforms multiplex information symbols in the DD domain, where each symbol is carried by an effective DD shaping pulse. Due to finite time and bandwidth resources, this effective pulse usually occupies a region larger than one ideal DD grid point, while conventional DD-domain signaling still keeps neighboring symbols approximately orthogonal. DD-domain FTN can be understood as reducing the spacing between DD-domain symbols, similar to reducing the time spacing from $T$ to $\delta T$ in conventional FTN. In this way, more symbols can be packed into the same DD resource region, so the excess occupancy of the effective DD shaping pulse is exploited to provide a higher symbol density. The cost is stronger non-orthogonal overlap among neighboring DD symbols, which requires more advanced DD-domain equalization. }

\subsection{FTN for Satellite Communications}

% Satellite communication and space networks have extraordinary potential to serve as the major force in the next generation of communication technologies, satellites operate on higher frequency bands which brings enormous amounts of bandwidth resources. However, the transmission power on a satellite is limited, the power amplifier on a satellite operates in a limited manner. Therefore, the constellation size used in satellite communications will be constrained. FTN allows a rate increase without consuming extra power, as $SNR_{tx}$ is fixed, the physical transmission is fixed as well. We can see that with low modulation order, we can achieve the same performance as high order modulation with FTN. 

Satellite communication and space networks hold immense potential as key enablers of next-generation communication technologies. Operating at higher frequency bands, satellites provide vast bandwidth resources, making them critical for global connectivity. However, satellite-based transmissions are constrained by limited power availability, and satellite power amplifiers must operate within strict efficiency limits. As a result, the modulation order in satellite communications is often restricted to maintain power efficiency. FTN presents a promising solution by enabling a higher transmission rate without requiring additional power. When \( \mathsf{SNR_{tx}} \) remains fixed, the physical transmission power remains unchanged. This allows FTN to achieve performance levels comparable to high-order modulation schemes while using lower modulation orders, making it particularly advantageous for power-constrained satellite communication systems.

\section{Conclusion}
{FTN signaling has evolved from a theoretical concept into a promising
candidate for improving spectral efficiency in future wireless systems.} We understand that FTN can deliver strong performance gains, especially in scenarios where bandwidth is tight or power is limited. Still, there are important issues to be resolved before FTN can be widely adopted. These include making equalization and decoding simpler, handling timing and synchronization challenges, and building hardware that can keep up. Encouragingly, recent advances in low-complexity detection and coding suggest that these challenges are within reach. As we push toward more efficient and flexible wireless systems, FTN is no longer just a theoretical curiosity, it is a technology worth taking seriously, especially in areas like integrated sensing and communications or in satellite links.

\section*{Acknowledgments}
This work was funded in part by the Scientific and Technological Research Council of Turkey, TUBITAK, under grant 122E248, and in part by a Discovery Grant awarded by the Natural Sciences and Engineering Research Council of Canada (NSERC).

\bibliographystyle{IEEEtran}
\bibliography{main}

\begin{IEEEbiographynophoto}{Zichao Zhang} [M] (zichaozhang@cmail.carleton.ca) is a Ph.D.
student in the Department of Systems and Computer Engineering, Carleton University. His research interest is faster than
Nyquist signaling.
\end{IEEEbiographynophoto}
\begin{IEEEbiographynophoto}{Melda Yuksel} [SM] (ymelda@metu.edu.tr) is an Associate Professor at Middle East Technical University, Ankara, Turkey. Her research interests are in wireless communications, communication theory and information theory with a current focus on faster than Nyquist transmission for future wireless communication systems. 
\end{IEEEbiographynophoto}
\begin{IEEEbiographynophoto}{Shuangyang Li} [M] (shuangyang.li@tu-berlin.de) is a Marie Skłodowska-Curie Actions (MSCA) research fellow at Technical University of Berlin, Germany. His research interests include signal processing, channel coding, applied information theory, and their applications to communication systems, with a specific focus on waveform designs.
\end{IEEEbiographynophoto}
\begin{IEEEbiographynophoto}{Gökhan Güvensen}  [M] (guvensen@metu.edu.tr) is an Assistant Professor at Middle East Technical University, Ankara, Turkey. His research interests include the design of digital communication systems and statistical signal processing. 
\end{IEEEbiographynophoto}
\begin{IEEEbiographynophoto}{Halim Yanikomeroglu}
[F] (halim@sce.carleton.ca) is a Chancellor's Professor and the Founding
Director of Non-Terrestrial Networks Lab at Carleton University, ON,
Canada.
\end{IEEEbiographynophoto}

\end{document}